\documentstyle[prl,aps,twocolumn,psfig]{revtex}
\begin{document}
\draft 
\title{Composite Fermions in Modulated Structures:
Transport and Surface Acoustic Waves}
\author{Felix von Oppen,$^{1,2}$ Ady Stern,$^1$ and Bertrand I.\
  Halperin$^3$}
\address{$^1$ Department of Condensed Matter Physics, The Weizmann
  Institute of Science, Rehovot 76100, Israel\\
 $^2$ Max-Planck-Institut f\"ur Kernphysik, Postfach 103980, 69029
 Heidelberg, Germany\\
 $^3$ Physics Department, Harvard University, Cambridge, Massachusetts
 02138}

\date{\today}

\maketitle

\def\kf{k_{\mbox{\tiny{F}}}}              
\def\vf{v_{\mbox{\tiny{F}}}}
\def\rd{\rho_{\mbox{\tiny{D}}}}            
\def\cf{{\mbox{\tiny{CF}}}}
\def\saw{{\mbox{\tiny  SAW}}} 

\begin{abstract} 

Motivated by a recent experiment of Willett {\it et al.}\ [Phys.\ Rev.\
Lett.\ {\bf 78}, 4478 (1997)], we employ semiclassical
composite-fermion theory to study the effect of a periodic density
modulation on a quantum Hall system near Landau level filling factor
$\nu=1/2$. We show
that even a weak density modulation leads to
dramatic changes in surface-acoustic-wave (SAW) propagation, 
and propose an explanation for
several key features of the experimental observations.
We predict that properly arranged {\it dc}
transport measurements would show a structure similar to that seen in
SAW measurements.

\end{abstract}      
\bigskip       
\narrowtext


Experiments on  surface acoustic waves (SAW)
propagation  above a
two-dimensional              electron             gas           (2DEG)
\cite{Willett-general,Willett-half}  provided  strong  support for the
composite-fermion approach to the compressible quantum-Hall state near
filling  factor $\nu=1/2$  \cite{HLR,Book}.  Measurements of  both 
absorption  and    velocity  shift of  SAW's  probe   the
conductivity  of  the    2DEG  at finite  wavevector     and frequency
\cite{Willett-general}.    In   this way,    Willett   {\it  et   al.}\
\cite{Willett-half} observed that  the absorption (velocity  shift) of
the SAW as function of filling factor exhibits  a maximum (minimum) at
$\nu=1/2$, implying   a  maximum  in the  conductivity.    Exactly  at
$\nu=1/2$, the  conductivity is found to be  linear in  the wavevector
for wavelengths smaller   than the composite-fermion   mean free path,
in agreement with  composite-fermion
theory.

Recently, Willett  {\it et al.}\ observed a  striking  effect in SAW
measurements near  $\nu=1/2$   on  samples  whose  electronic  density
$n({\bf r})={\bar n}+\delta n(x)$ is   periodically modulated in {\it  one}
direction,  say $\hat x$ \cite{Willett-mod}.   When the SAW propagates
in the  ${\hat y}$--direction,  a rather weak  density modulation
($\delta  n/n\le0.05$) turns  the minimum   in  the velocity shift  at
$\nu=1/2$ into a surprisingly robust maximum:
Unlike the former, the magnitude and width of the latter
are almost  independent of the SAW  wavevector $q$ and  the modulation
period  $a$ (for  sufficiently   small $a$).  
In contrast,  the modulation has no significant
effect when the SAW propagates in the $\hat x$ direction.

In this Letter, we analyze $dc$ transport, SAW velocity shift and SAW
absorption  in
modulated systems near $\nu=1/2$. We employ semiclassical
composite-fermion theory \cite{HLR,SimonHalperin}, which
allows one to derive a Boltzmann equation for composite fermions
(CF's). Within this theory, one attaches two Chern-Simons flux quanta to
each electron.  The resulting
quasi-particles -- CF's -- experience an effectively
reduced magnetic field, ${\cal B}({\bf r})=B-(2h/e)n({\bf r})$ so that
a density modulation leads to a modulated magnetic field ${\cal
B}({\bf r})$.  Thus, experiments on modulated structures
test a fundamental aspect of the theory.

We first consider $dc$ transport in modulated systems with
very large modulation period $a$. For such systems the current
${\bf J}({\bf r})$ is related to the electric field ${\bf E}({\bf r})$
by a {\it local} resistivity tensor $\rho(x)$. 
The measurable quantity, however, is the
macroscopic resistivity $\rho^{\mbox{\tiny mac}}$, relating the
{\it spatially averaged} current and field. We now show that the 
modulation makes  
$\rho^{\mbox{\tiny mac}}$ anisotropic although
locally $\rho_{xx}=\rho_{yy}$.
The local resistivity
is a function of the local density $n(x)$ and can be written
as $\rho(x)\equiv {\bar\rho}+\delta\rho(x)$, where $\delta\rho$ has
zero spatial average. [Here and below, bars denote quantities 
in the unmodulated system]. We assume a strong magnetic
field so that $\rho_{xy}\gg \rho_{xx},\rho_{yy}$, and neglect the
diagonal elements of $\delta\rho$. 
Since the density depends only on $x$, the current,
electric field, and $\rho$ are all independent of $y$.
Conservation of current then implies that $J_x$ is
uniform in space. From Maxwell's equations, we have 
${\bf \nabla}\times {\bf E}=0$, which implies
$\delta\rho_{yx}J_x+\rho_{yy}\delta J_y=0$. Here,
$\delta J_y$ is the modulated part of the current in the $y$ 
direction, whose spatial average is zero.  Thus \cite{Gerhardts}, while
$\rho^{\mbox{\tiny mac}}_{yy}={\bar\rho}_{yy}$, 
\begin{equation}
\rho^{\mbox{\tiny
    mac}}_{xx}={\bar\rho}_{xx} (1 + \beta^2 \delta n_{rms}^2 / {\bar
n}^2), \label{resfor} \end{equation} 
where $\delta n_{rms} $ is the
root-mean-square value of $\delta n$, the deviation of the local
electron density from its mean value $\bar n$, and we have defined
$\beta$ by, 
\begin{equation} 
\delta \rho_{xy} \equiv \beta\rho_{xx}
\delta n /{\bar n} .  \label{bih3} \end{equation} 
In a naive Drude
picture $\beta=\rho_{xy}/\rho_{xx}\gg 1$.  In the quantum Hall regime,
empirical observations \cite{Stormer} that $\rho_{xx}$ is proportional
to the derivative of $\rho_{xy}$ with respect to the logarithm of the
magnetic field (``resistivity law''), together with the observation
that $\rho_{xy}$ is primarily determined by the filling factor $\nu$,
suggest that the coefficient $\beta$ is in fact a constant,
independent of the applied magnetic field, and weakly temperature
dependent, of order 20 or more for high-mobility samples.  Thus, {\it
a weak density modulation, while having no effect on
$\rho^{\mbox{\tiny mac}}_{yy}$,  strongly enhances
$\rho^{\mbox{\tiny mac}}_{xx}$}.

An SAW transmitted above a 2DEG gives rise to a ``bare" electric field
${\bf E}^\saw_{{\bf q},\omega}$, parallel to $\bf q$, due to the
piezoelectric effect in GaAs.  In unmodulated systems, the screening
response of the 2DEG leads to an absorption and a velocity shift of
the SAW, proportional to the imaginary and real parts, respectively,
of $(1 + i \sigma_{\alpha\alpha} / \sigma_m )^{-1} $, where $\sigma$
is the electronic conductivity at wavevector $\bf q$ and frequency
$\omega=v_sq$ ($v_s$ being the sound velocity), $\alpha$ is the
direction of ${\bf q}$, and $\sigma_m = \epsilon v_s/2 \pi$, with
$\epsilon$ the appropriate background dielectric
constant\cite{Willett-general,SAWtheory}.  As customary, the
velocity shift is given relative to its value for
$\sigma_{\alpha\alpha}=\infty$.  One may naively conjecture that in a
modulated system $\sigma$ should be replaced by $\sigma^{\mbox{\tiny
    mac}}$.  Since $\sigma^{\mbox{\tiny mac}}_{yy}\approx
\rho^{\mbox{\tiny mac}}_{xx} / ({\bar\rho}_{xy})^2$, such a conjecture
(to be partially verified below) predicts a modulation--induced
suppression of velocity shift when ${\bf q}||{\hat y}$, and no effect
when ${\bf q}||{\hat x}$.  In strong magnetic fields $\sigma_m$ is
large compared to $ \sigma_{\alpha \alpha}$, so that a decrease in the
velocity shift corresponds to an increase in the absorption and vice
versa.

The local approximation obviously applies when 
the modulation period $a$ is much larger than the
composite fermion scattering length $\ell_{\rm tr}$.  
Even when this condition
does not hold the resistivity $\rho$ is still local far from
$\nu=1/2$, where the composite
fermion cyclotron radius is much smaller than $a$. Thus, away from
$\nu=1/2$ we still expect the
modulation  to enhance the $dc$ resistivity following Eq.~(\ref{resfor})
and similarly to enhance SAW absorption and suppress the velocity
shift.
{\it  Near $\nu=1/2$, however, we expect these effects to be greatly 
reduced
if $a <  \ell_{\rm tr}$,
leading to a local minimum in the resistivity and SAW absorption (as a
function of magnetic field), and a
maximum in the SAW velocity shift.} \cite{Footnote} 
The detailed analysis 
we present below
confirms these general expectations.

We now proceed to  study SAW propagation without assuming a local
resistivity tensor.  The electric field ${\bf
   E}^{\mbox{\tiny SAW}}_{{\bf q},\omega}$ induces electronic currents
 and densities in the 2DEG.  In linear response, the current ${\bf J}$
 is related to the SAW field by ${\bf E}^{\mbox{\tiny SAW}}=
 \rho_\saw{\bf J}$.  The matrix $\rho_\saw$ differs from the
 electronic resistivity, $\rho$, which relates the current to the
 total electric field ${\bf E}^\saw+{\bf E}^{\rm ind}$
acting on the electrons.  The field ${\bf E}^{\rm ind}$
due to the induced electronic charge density
can be linearly related to the
current by ${\bf E}^{\rm ind}={\cal U}{\bf J}$ (with ${\cal U}$ 
given below).  Clearly, $\rho_\saw=\rho-{\cal U}$.
In modulated systems,  both  $\rho$ and $\rho_\saw$  are 
non-diagonal 
in momentum: an electric  field of wavevector $\bf q$ induces
currents  and densities of wavevectors ${\bf  q}+l{\bf p}$,
where $l$ is an integer and ${\bf p} = 2 \pi \hat{x} /a$.  
We   use     the      notation   $\left(\rho_{\saw}({\bf
    q})\right)_{\alpha\beta}^{kl}$   for the    ratio    of the  field
$E^{\saw}_\alpha$  of  wavevector ${\bf  q}+k{\bf  p}$  to an  applied
current  $J_\beta$ of  wavevector  ${\bf  q}+l{\bf p}$.  The  $\omega$
dependence   is left implicit since   all  quantities are diagonal  in
$\omega$.  Without modulation,  $\rho$, $\rho_\saw$, and their 
inverses
are diagonal in  the modulation indices $j,l$.   By Coulomb's  law and
the   continuity equation,  ${\cal      U}_{\alpha\beta}^{jl}=
-i\delta_{jl}\ 
\frac{(q_\alpha+lp_\alpha)(q_\beta+lp_\beta)}{\omega}\frac{2\pi
  }{\epsilon |{\bf q}+l{\bf p}|}$.

The rate of energy absorption by the electrons, $P$, is
\begin{equation} 
P={\rm Re}\left[{\bf
J}^*\cdot({\bf E}^{\saw}+{\bf E}^{\rm ind})\right]= 
{\rm Re}\left({\sigma_{\saw}}\right)^{00}_{\alpha\alpha}|E^\saw|^2 
\label{absorb}
\end{equation}
with   $\sigma_{\saw}=\rho_{\saw}^{-1}$.   The
SAW velocity shift is  proportional to  ${\rm Im}
\left(\sigma_\saw\right)^{00}_{\alpha\alpha}+\sigma_m$
\cite{Willett-half,SAWtheory}. 
A related formalism was independently developed
in Ref.\ \cite{Levinson}. 

Our   starting  point  for  calculating  $\sigma_{\saw}$ is a
Boltzmann equation   describing the   semiclassical  dynamics  of 
CF's      in       a     modulated         potential
\cite{SimonHalperin}.      This        equation
incorporates the  coupling  of  the  CF's to  the  Chern-Simons fields
describing the  interactions of   the  charges with  the  attached 
flux  quanta.  It is  valid 
close  to $\nu=1/2$  where   the  quantum mechanics  of   CF's can  be
neglected. (All relevant length scales are assumed large compared to 
$1/k_F$.)
The Boltzmann equation is   an equation for $\delta  n_{\bf
  p}({\bf r},t)$,  the deviation of  the composite-fermion phase-space
distribution   function from  its  equilibrium  value.   Within linear
response,
\begin{eqnarray}
  &&\left\{\partial_t+{\bf v_p\cdot\nabla_{\!r}}-({\bf \nabla_{\!\bf r}}
  V^{\rm sc})\cdot{\bf\nabla}_{\!{\bf p}}+e[{\bf v_p}\times{\bf\hat
    z}{\cal B}]\cdot\nabla_{\!\bf p}\right\}\delta n_{\bf  p}
   \nonumber\\
   &&\,\,\,\,\,\,\,\,\,\,\,\,\,\,\,
  +e{\bf {\cal E}}\cdot
  \nabla_{\!\bf p}n_{\bf  p}^{(0)}-{\cal I}[\delta n_{\bf  
  p}-\sum_{\bf p'}\,\delta n_{\bf p'}]=0.
\label{boltz} 
\end{eqnarray}
Here $V^{\rm sc}(x)$ is the self-consistent equilibrium electrostatic
potential creating the modulation, ${\bf v_p}$ is the velocity of a
composite fermion of momentum ${\bf p}$ and $\cal I$ is the impurity
scattering collision integral.  The equilibrium value of the
phase space distribution function is the
Fermi-Dirac distribution $n_{\bf p}^{(0)}({\bf
  r})=f_\mu(p^2/2m+V^{\rm sc}(x))$  with a chemical potential $\mu$.    The composite fermions are subject to a spatially
modulated effective magnetic field ${\cal B}(x)$. The effective
electric field ${\bf {\cal E}}={\bf E}^{\saw}+{\bf E}^{\rm ind}+{\bf
  E}^{\rm CS}$ is composed of the physical field ${\bf E}^{\saw}+{\bf
  E}^{\rm ind}$, and the Chern--Simons electric field ${\bf E}^{\rm
  CS}=(2h/e^2){\bf J}\times{\bf\hat z}$.  The modulation enters Eq.\ 
(\ref{boltz}) through $V^{\rm sc}(x)$, ${\cal B}(x)$, and the Fermi
velocity. The electronic current induced by ${\bf {\cal E}}$ is ${\bf
  J}({\bf r},t)=\sum_{\bf p}{\bf v_p}\delta n({\bf p},{\bf r},t)$.  We
emphasize that while we use  composite-fermion theory, we present
here only  measurable {\it electronic} response functions.

The essential physics is captured by a perturbative calculation of
$\sigma_{\saw}$ to second order in the density modulation $\delta n$.
In this calculation we consider long SAW wavelengths
$q\ell_{\rm tr}\ll 1$, weak disorder $\kf\ell_{\rm tr}\gg 1$ and $p\gg
q$.  We first consider the SAW wavevector  to be ${\bf q}=q{\hat y}$,
i.e.,  perpendicular
to the modulation wavevector. We write
$\rho_\saw={\bar\rho}_\saw+\delta\rho$ and
$\sigma_\saw={\bar\sigma}_\saw+\delta\sigma$. Since the Boltzmann
equation is a convenient tool for calculating $\delta\rho$ in powers
of $\delta n$, we write
\begin{equation}
  (\delta\sigma_{\saw})^{00}_{yy}\simeq
  \left[{\bar\sigma}_{\saw}\left(-\delta\rho+\delta\rho\,
  {\bar\sigma}_{\saw}\,\delta\rho\right){\bar\sigma}_{\saw}
  \right]_{yy}^{00}.
\label{orders}
\end{equation} 
   Since  ${\bar\sigma}$   is diagonal in    its
superscripts,    both the    rightmost    and leftmost   matrices  are
$({\bar\sigma}_\saw)^{00}$. We find  $({\bar\sigma}_\saw)^{00}$ using
${\bar\sigma}_\saw=({\bar\rho}-{\cal U})^{-1}$. 
The $xx$ element does not contribute to (\ref{orders}). 
The
off-diagonal elements  are larger than  the $yy$
element by a factor  $\kf\ell_{\rm tr}$, and are given by
$\pm\frac{{\bar\sigma}_{xy}({\bf q})}{1+i{\bar\sigma}_{yy}/\sigma_m}$. 
Since $q\ell_{\rm tr}\ll 1$,  we may approximate ${\bar\sigma}_{xy}({\bf
  q})$ by   its $q=0$ value.  The biggest contribution to
the   first   term   in (\ref{orders}),    then,    is proportional to
$\delta\rho^{00}_{xx}$,  which   is    what  is   measured in
Weiss-oscillation measurements \cite{Gerhardts}. Its contribution here
is smaller by a factor $\kf\ell_{\rm tr}$ than that of the second term
in (\ref{orders}).

For the second term of (\ref{orders}) we use the Boltzmann equation 
(\ref{boltz}) 
to calculate $\delta\rho$ 
to first order in $\delta n$.  We find that as long as $\kf a\gg 1$
(a condition well satisfied by the experimental system), the Hall 
components of
$\delta\rho^{l0}$ are
\begin{equation}
      \delta\rho_{xy}^{l0}=-\delta\rho_{yx}^{l0}=\bar\rho_{xy}
       \frac{\delta n_l}{n}
\label{deltarhojh}
\end{equation}
with $\delta n_l=(1/a)\int_0^a dx\delta n(x)\exp{(i2\pi lx/a)}$.
The diagonal components $\delta\rho_{xx}^{l0}, \delta\rho_{yy}^{l0}$
are smaller by a factor of $\kf \ell_{\rm tr}$. 
By   the Onsager  symmetry,
$\rho^{0l}_{\alpha\beta}(B)=\rho^{l0}_{\beta\alpha}(-B)$
\cite{LandauLifshitz}.  These expressions, which are 
obvious in the local limit $a\gg \ell_{\rm tr}$, hold irrespective of 
the ratio of $a$ to $\ell_{\rm tr}$.

Eqs.~(\ref{orders},\ref{deltarhojh}) and the expression for 
${\bar\sigma}^{00}_{xy}$
suggest that the second term in (\ref{orders}) can be approximated by,
\begin{equation}
(\delta \sigma_{\rm  SAW})^{00}_{yy}\simeq
\sum_{l\ne0}
      ({\bar\sigma}_{\saw})^{00}_{yx}\delta\rho^{0l}_{xy}
({\bar\sigma}_{\saw})^{ll}_{yy}\delta\rho^{l0}_{yx}
({\bar\sigma}_{\saw})^{00}_{xy}
\label{explicit}
\end{equation}
This     is   indeed   the   case,   since      we now   show     that
$({\bar\sigma}_{\saw})^{ll}$ is dominated by its $yy$ element.

The response function $(\bar\sigma_{\saw})^{ll}$ relates an externally
applied  electric field of wavevector ${\bf  q}+l{\bf p}$ to a current
of   the same wavevector  {\it  in  an unmodulated  system}.  With the
expression for ${\cal U}$, one finds for the  inverse of
$(\bar\sigma_{\saw})^{ll}$, 
\begin{equation}
  ({\bar\rho}_\saw)^{ll}=
  \left[\begin{array}{cc}
  {\bar\rho}_{xx}(l{\bf p})+i\frac{lp}{q\sigma_m}  
  & {\bar\rho}_{xy}(l{\bf p})+\frac{i}{\sigma_m}\\
  {\bar\rho}_{yx}(l{\bf p})+\frac{i}{\sigma_m}
  &{\bar\rho}_{yy}(l{\bf p})+i\frac{q}{lp\sigma_m}
  \end{array}\right]
\label{oneone}
\end{equation}
Here we  approximated ${\bf q}+l{\bf  p}\simeq l{\bf p}$.  Since $p\gg
q$  and      $\sigma_m\ll   e^2/h$,     the   biggest   element     in
$({\bar\rho}_\saw)^{ll}$ is the $xx$ component, and, consequently, its
inverse is dominated by  the $yy$ component. If $q$ is small enough 
such that
$(4h/e^2)\sigma_m\ll p^2/q\kf$ and $(pv_s)/(q\vf)\gg 1$ (with $\vf$ the 
composite fermion Fermi velocity),  
then $ (\bar\sigma_{\saw})^{ll}_{yy}\approx 1/{\bar\rho}_{yy}(l{\bf
  p})\approx(e^2/2h)^2/{\bar\sigma}_{xx}(l{\bf p})$.  The largeness of
the $yy$ element, $(\bar\sigma_{\saw})^{ll}_{yy}$, which plays an
important role in our calculation, is in marked contrast to the
conductivity matrix ${\bar\sigma}^{ll}$, whose largest elements are
the off-diagonal ones, due to the strong magnetic field.  This
contrast reflects the fact that in the modulated system, the SAW field
${\bf E}^\saw$ in the $\hat y$ direction is accompanied by a large
induced field in the $\hat x$ direction.  The current is almost
perpendicular to the {\it total} electric field and hence has a
sizable component in the $\hat y$--direction.

Finally, using (\ref{explicit}) and (\ref{oneone}), we find
\begin{equation}
  (\delta\sigma_{\saw})^{00}_{yy}\simeq\sum_{l\ne0}\left({\delta
      n_l\over n}{e^2/2h\over1+i\bar\sigma_{yy}/\sigma_m}\right)^2
      {1\over\bar\sigma_{xx}(l{\bf p})}.
\label{resultme}
\end{equation}
This is the central result of our analytical calculation.  In the
local limit, $p\ll 1/\ell_{\rm tr}$, this expression can be obtained
from the analysis described above Eq.\ (\ref{resfor}).  An analogous
calculation shows that the effect for SAW propagation parallel to the
modulation direction, $q=q{\hat x}$, is smaller by a factor of order
$(\kf\ell_{\rm tr})^2\sim 10^{3}-10^{4}$. Experimentally, indeed, the
modulation has no observable effect in this case.  The 
modulation contribution to the macroscopic conductivity is given by
a similar analysis with $\sigma_m=\infty$.

\begin{figure}
\centerline{\psfig{figure=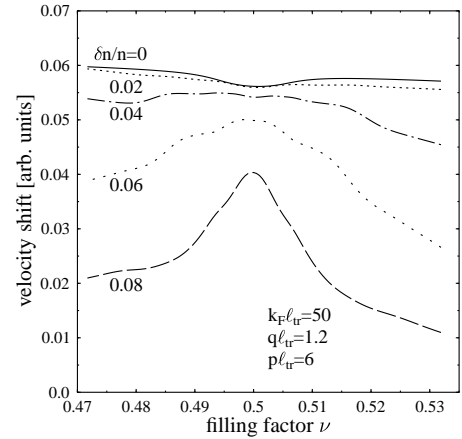,width=7.3cm}}
\caption{SAW velocity shift vs filling factor
 for different modulation strengths.
  The SAW wavevector is $q\ell_{\rm tr}=1.2$ and the
  modulation  wavevector $p\ell_{\rm   tr}=6$.    With
  increasing modulation, the minimum in velocity shift  changes  into
  a maximum, consistent with the experimental
  results of Ref.~\protect\cite{Willett-mod}.}
\end{figure}

We now show that the predictions of Eq.\ (\ref{resultme}) are in 
qualitative agreement with key experimental results. The influence of the
modulation is essentially determined by $\bar\sigma_{xx}(l{\bf p})$.
This quantity is known, both from previous experiments
\cite{Willett-half} and from theory \cite{HLR,Book}, to exhibit
a maximum as a function of magnetic field at $\nu=1/2$ when $l|{\bf
  p}|\gg1/\ell_{\rm tr}$.  Thus Eq.~(\ref{resultme}) predicts that the
modulation-dependent contribution to the absorption (velocity shift)
has a minimum (maximum) around $\nu=1/2$, once the modulation period
$a$ is smaller than $\ell_{\rm tr}$.  The trends of the
modulation-independent contribution, ${\bar\sigma}_{yy}^{00}$, around
$\nu=1/2$ are just opposite.  For strong enough modulation, it is the
modulation--dependent contribution which determines the type of
extremum points at $\nu=1/2$, {\it in agreement with the new effect
  observed by Willett et al.}

The analytical results are well supported by our numerical solutions
of the Boltzmann equation, which are not restricted to weak modulation
or to the regime $q\ll p,\ell_{\rm tr}^{-1}$.  Numerically, we
directly compute the response to the applied SAW field. We restrict
ourselves to a modulated magnetic field with a single Fourier
component, and employ the isotropic relaxation-time approximation
\cite{Mirlin} to account for impurity scattering.  Representative
results for the SAW velocity shift as function of filling factor around
$\nu=1/2$ are shown in Fig.~1.  At zero modulation, the velocity shift
exhibits the usual minimum at $\nu=1/2$.  As the modulation is
increased, the minimum disappears and a maximum develops in accord
with the analytical conclusions above.  The effect of the modulation
gets stronger as one gets farther from $\nu=1/2$. As seen in Fig.~1,
the modulation induced peak in the velocity shift is more pronounced
than the minimum observed at zero modulation, in
agreement with the experiment.  We emphasize that
we found similar behavior over a wide range of the parameters $q$ and
$p$.  

Experimentally, the modulation-induced peak in velocity shift was
strikingly insensitive to the SAW wavevector $q$ and the modulation
wavevector $p$. Fig.\ 2 shows our results for that peak for realistic
values of $q$ and $p$.  Clearly, the width and magnitude of the peak
are rather stable over a substantial parameter range -- a factor 2 in
modulation period and a factor 3 in SAW wavelength, in good
qualitative agreement with the experiment.

There is a point of disagreement between our theory and the
experiment.  Theoretically, the maximum in the velocity shift is
primarily due to a decrease in the velocity shift away from $\nu=
1/2$, rather than to an increase in its value at $\nu = 1/2$.
Experimentally, there seems also to be a sharp increase in
$\sigma^{00}_{yy}$ at $\nu=1/2$.  This may be caused by unknown
physical effects that were omitted from our model. However,
another conceivable explanation for this difference within our model
might be that the material parameter $\sigma_m$ increases with the
voltage applying the modulation, leading to an absolute increase in
the velocity shift at $\nu=1/2$.

\begin{figure}
\centerline{\psfig{figure=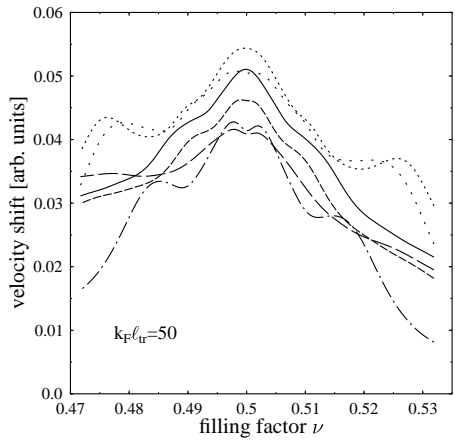,width=7.3cm}}
\caption{Modulation-induced maximum in the SAW velocity shift vs
  filling factor for  different SAW  
  and modulation wavevectors $q$ and  $p$. The parameters ($q\ell_{\rm tr}$,
  $p\ell_{\rm tr}$, $\delta n/n$) are for the full line (0.6, 6, 0.06),
  short dotted (0.6, 10, 0.07), 
  short dashed (1.2, 6, 0.07),
  wide dotted (1.2, 10, 0.08),
  long dashed (2, 6, 0.08), and  
  dash-dotted (2, 10, 0.11).}
\end{figure}

In fact, the values of $\sigma_m$ which were used  
to relate SAW propagation to
$\sigma$ in unmodulated samples have been much larger
than the theoretical values, for
reasons which are not understood \cite{HLR}.  Thus a dependence on the 
gate voltage, and perhaps on the direction of SAW propagation, is not
inconceivable.  An increase in the mean free path $\ell_{\rm tr}$ for
increasing gate voltage in the modulated samples would also give
qualitatively similar effects.

Our calculation for $dc$ transport, whose result was stated below
Eq.\ (\ref{resultme}), predicts anisotropies in the macroscopic $dc$
conductivity and resistivity tensors. Specifically, we find that 
both $\rho^{\mbox{\tiny mac}}_{xx}$ and $\sigma^{\mbox{\tiny mac}}_{yy}$
exhibit minima near $\nu=1/2$ (with the modulation in the $x$ direction),
observable in Hall-bar and Corbino geometry, respectively. Transport
experiments reported in Ref. \cite{Willett-mod} have not shown these
effects.  However, it is not clear what are the actual current paths
in this experiment.  Very recent experimental results of Smet {\it
et al.}\ \cite{Smet} are in qualitative agreement with our theory.

In conclusion, we find within a semiclassical composite-fermion
approach that a weak density modulation can dramatically affect both
$dc$ transport properties and SAW propagation near $\nu=1/2$. Our
results are in agreement with many key features of the  experimental
results. Details of the calculation will be published elsewhere. 

We acknowledge financial support from the US-Israel Binational
Science Foundation (AS and BIH, 95-250/1), the Minerva foundation (AS
and FvO), and NSF grant DMR94-16190 (BIH).  We thank
B.\ Elattari for collaboration in early stages of this work,
O. Entin, Y.  Levinson, J. Smet, and R.L.  Willett for 
instructive discussions of unpublished results.


\begin{thebibliography}{10}

\bibitem{Willett-general} R.L.\ Willett, Surf.\ Sci.\ {\bf 305},
76 (1994).

\bibitem{Willett-half} R.L.\ Willett, R.R.\ Ruel, K.W.\ West, and 
L.N.\ Pfeiffer, {\bf 71}, 3846 (1993). 

\bibitem{HLR} B.I.
  Halperin, P.A.\ Lee, and N.\ Read, Phys.\ Rev.\ B {\bf 47}, 7312
  (1993).  

\bibitem{Book} B.I.\ Halperin, in {\em Novel Quantum
    Liquids in Low-Dimensional Semiconductor Structures}, edited by S.\
  Das Sarma and A.\ Pinczuk (Wiley, New York, 1996).

\bibitem{Willett-mod} R.L.\ Willett, K.W.\ West, and L.N.\ Pfeiffer, Phys.\
  Rev.\ Lett.\ {\bf 78}, 4478 (1997). 
  
\bibitem{SimonHalperin} S.H.\ Simon and B.I.\ Halperin, Phys.\ Rev.\ B
  {\bf 48}, 17368 (1993);
A.\ Stern and B.I.\ Halperin, Phys.\ Rev.\ B
  {\bf 52}, 5890 (1995).

\bibitem{Gerhardts} R.\ Menne and R.R.\ Gerhardts,
cond-mat/9709072.

\bibitem{Stormer} H. L. Stormer {\it et
    al.}, Sol.\ Stat.\ Comm.\ {\bf 84}, 95 (1992). S.H.\ Simon and
  B.I.\ Halperin, Phys.\ Rev.\ Lett.\ {\bf 73}, 3278 (1994).

\bibitem{Footnote} By contrast, when the modulation becomes {\it too}
  strong, the local composite fermion cyclotron radius will be small
  (compared to $a$) even when the average filling factor is $1/2$.
  Then, the conductivity is local even at $\nu=1/2$ and the structure
  at $\nu=1/2$ should disappear. This is consistent with the
  obervations of Ref.\ \cite{Willett-mod}.

\bibitem{SAWtheory} S.\ Simon, Phys.\ Rev.\ B {\bf 54}, 13878 (1996).

\bibitem{Levinson} Y.\ Levinson {\it et al.},
private communication and cond-mat/9712276.
  
\bibitem{Mirlin} Our numerical results are largely insensitive to the 
issues raised in  A.\ Mirlin and   P.\ W\"olfle, Phys.\ Rev.\ Lett.\ 
  {\bf 78}, 3717 (1997). 

\bibitem{LandauLifshitz} L.D.\ Landau and E.M.\ Lifshitz {\it Statistical
    Physics}, sec.\ 122, Pergamon Press (1978). 

\bibitem{Smet} J.\ Smet, K.\ von Klitzing, D.\ Weiss, and W.\ Wegscheider,
(to be published). 

\end{thebibliography}
\end{document}